\documentclass[twocolumn,prb,showpacs,superscriptaddress]{revtex4}
\usepackage{graphicx}
\usepackage{dcolumn}
\usepackage{bm}

\begin{document}

\title{Impurity effect as a probe of the paring symmetry in $\mathrm{BiS_{2}}$-based superconductors}

\author{S. L. Liu}
\affiliation{College of Science, Nanjing
University of Posts and Telecommunications,\\
Nanjing 210003, China}

\date{\today}

\begin{abstract}

The impurity effects are studied for the
$\mathrm{BiS_{2}}$-layered superconductors based on a two-orbital
model with the Bogoliubov-de-Gennes technique. The superconducting
critical temperature ($\mathrm{T_{c}}$) is calculated as a
function of the impurity concentration. Significant reduction of
$\mathrm{T_{c}}$ is found for the spin singlet nearest-neighboring
paring d-wave state and the spin triplet next-nearest-neighboring
(NNN) paring p-wave state, but no depression of $\mathrm{T_{c}}$
for isotropic s-wave state. The single impurity effects for
various paring states are also explored. The impurity resonance
peak in the local density of states spectrum is found only for the
d-wave state. For the spin triplet NNN paring p-wave state, two
in-gap peaks occur in the case of positive impurity potential and
a single in-gap peak is found for the negative impurity potential,
and it shifts towards the lower energy with the potential strength
decreasing. These results can be used to detect the paring
symmetry of the $\mathrm{BiS_{2}}$-based superconductors.

\end{abstract}

\pacs{74.70.Xa, 74.55.+v}

\maketitle

Superconductors with layered crystal structures such as
cuprates\cite{Pickett1989}, $\mathrm{Sr_{2}RuO_{4}}$\cite{Maeno},
$\mathrm{MgB_{2}}$\cite{Nagamatsu2001},
$\mathrm{Na_{x}Co_{2}O_{2}}$\cite{Takada} and iron
pnictides\cite{Kamihara2008} have generated enormous research
interest. The low dimensionality can affect the electronic
structure and can realize high transition temperatures and/or
unconventional superconductivity mechanisms. Quite recently,
superconductivity has been observed in the
$\mathrm{BiS_{2}}$-based compounds\cite{Mizuguchi}, such as
$\mathrm{Bi_{4}O_{4}S_{3}}$\cite{Mizuguchi},
$\mathrm{LaO_{1-x}F_{x}BiS_{2}}$\cite{Mizuguchib},
$\mathrm{NdO_{1-x}F_{x}BiS_{2}}$\cite{Demura},
$\mathrm{CeO_{1-x}F_{x}BiS_{2}}$\cite{Xing}, and
$\mathrm{PrO_{0.5}F_{0.5}BiS_{2}}$\cite{Jhab},
 where the
$\mathrm{BiS_{2}}$-layer is a basic unit. Soon after that,
considerable efforts have been paid on the investigation of the
physical properties in this
system\cite{Usui,Sheng,Tan,Singh,Awana,Kotegawa,Zhou,Wan,Takatsu,Sathish,Jha,Lei,Tanb,Deguchi9,Li}.
The discovery of the new basic superconducting (SC) layers may
open new fields in physics and chemistry of low-dimensional
superconductors.

The $\mathrm{BiS_{2}}$-based materials compose of the stacking of
$\mathrm{BiS_{2}}$-SC layers and the spacer layers, which
resembles those of high-$\mathrm{T_{c}}$ cuprates and the Fe-based
superconductors. The first principle band structure
calculations\cite{Usui,Mizuguchi} predict that the dominating
bands for the electron conduction as well as for the
superconductivity are derived from the Bi $6p_{x}$ and 6$p_{y}$
orbits. According to the Hall effect and the magnetoresistance
measurements\cite{Sheng}, an exotic multi-band feature is found in
the $\mathrm{Bi_{4}O_{4}S_{3}}$ compounds and the SC pairing
occurs in the one-dimensional chains. Electrical resistivity
measurements under pressure\cite{Kotegawa} reveal that
$\mathrm{Bi_{4}O_{4}S_{3}}$ and $\mathrm{LaO_{1-x}F_{x}BiS_{2}}$
have different $T_{c}$ versus pressure behavior, and the Fermi
surface of $\mathrm{LaO_{1-x}F_{x}BiS_{2}}$ may be located in the
vicinity of some band edges, which leads to the instability for
superconductivity. It is found that the parent phase is a bad
metal in the $\mathrm{CeO_{1-x}F_{x}BiS_{2}}$ system\cite{Xing},
while a band insulator in the case of $\mathrm{Bi_{4}O_{4}S_{3}}$
compounds\cite{Mizuguchi}. By doping electrons into the system,
superconductivity appears together with an insulating normal
state\cite{Xing}, which is obviously different from that of the
cuprates and the iron pnictides.

The most fundamental issue in this $\mathrm{BiS_{2}}$-based
compound is the superconductivity mechanism, which is still
unclear. It is predicted that the electron-phonon coupling
constant is as large as $\lambda=0.85$, which indicates that
$\mathrm{LaO_{1-x}F_{x}BiS_{2}}$ is a strongly coupled
electron-phonon superconductor\cite{Wan,Li}. However, because
electron-electron correlation generally is more important in a low
dimensional system, the correlation effect might play an important
role in driving superconductivity even if the $p$-orbits of Bi are
much less localized compared with the $d$-orbits in cuprates and
iron-based superconductors\cite{Liang2012}. Experimentally, no
magnetically ordered phase has been detected so far in the
$\mathrm{BiS_{2}}$ compounds. This apparent absence of magnetism
in the $\mathrm{BiS_{2}}$ compounds may still locate them in the
same category as LiFeAs, FeSe, and possibly
$\mathrm{Sr_{2}VO_{3}FeAs}$, which are also non magnetic but their
pairing properties are widely believed to still be originated from
the short-range magnetic fluctuations\cite{Martins2012}. Moreover,
a good nesting of the Fermi surface at wave vector
$\mathbf{k}=(\pi,\pi,0)$ has been found\cite{Usui,Wan}. It is also
proposed that the SC pairing is strong and it exceeds the limit of
the phonon mediated picture\cite{Sheng}. The anharmonic model
calculation shows that the vicinity of the charge-density-wave
instability is essential for the superconductivity\cite{Wan}. It
is reminiscent to the proximity to spin-density-wave of iron
pnictides, which is well established\cite{Kamihara2008}. The
correlation effect, therefore, seems to be a good candidate
responsible for the SC pairing in these materials. Thus, the spin
fluctuation is also proposed to account for the SC pairing in this
family\cite{Zhou,Liang2012,Martins2012}.

Another puzzle is the SC paring symmetry of the
$\mathrm{BiS_{2}}$-based superconductors. Due to the large
electron-phonon coupling constant, the paring symmetry may usually
be a conventional s-wave state with isotropic SC gap. However, the
correlation effect is also proposed to account for the SC paring,
which may lead to the paring symmetry in a complicated situation
in these compounds. Moreover, based on the the first principle
calculation\cite{Usui}, it is found that the Fermi level for the
nominal composition is located in the vicinity of the topological
change in the Fermi surface. This gives a possibility that the SC
symmetry changes depending on the doping level, provided that the
superconductivity originates from the electric correlation.
Considering various many-body interactions, possible paring states
are proposed, such as sign conserving/reversing s-wave and d-wave
states\cite{Usui,Zhou}. Furthermore, the first principle
calculation indicates the quasi-one-dimensional bands\cite{Usui},
which may lead to the spin-triplet pairing. In the
$\mathrm{CeO_{1-x}F_{x}BiS_{2}}$ compounds, superconductivity and
ferromagnetism (FM) are found to coexist\cite{Xing}, where FM may
arise from the Ce moments. The coexistence of superconductivity
and FM challenges the spin-singlet paring mechanism, which reminds
us of the $\mathrm{Sr_{2}RuO_{4}}$ material, whose short-range
ferromagnetic spin fluctuations give rise to the triplet pairing
with p-like symmetry\cite{Miyake1999}.

The impurity effect is one of the most important tools for
identifying the nature of the paring state and the microscopic
properties, which has been successfully carried out in both
conventional superconductors\cite{Woolf1965} and unconventional
ones, such as cuprates\cite{Balatsky2006} and iron
pnictides\cite{Zhang2009,Zhou2011,zhou20112,Jiang2011,Bang2009,Vorontsov2009,Liu2012}.
In this paper, we will study theoretically the impurity effect in
the $\mathrm{BiS_{2}}$-based materials and look into the pairing
symmetry based on the two-orbital model\cite{Usui} and the Bogoliubov-de-Gennes (BdG)
equations. The impurity concentration dependence of the critical
temperature is explored. Significant reduction of $\mathrm{T_{c}}$
is found for the spin singlet nearest-neighboring (NN) paring d-wave
state and the spin triplet next-nearest-neighboring (NNN) paring
p-wave state, while no depression of $\mathrm{T_{c}}$ for
isotropic s-wave state. The single impurity effects for various
paring states are also calculated. The impurity resonance peak in
the local density of states (LDOS) spectrum is found only for the
d-wave state. The evolution of the resonance peak with the
impurity strength is calculated as well. For the spin triplet NNN
paring p-wave state, two in-gap peaks occur in the case of
positive impurity potential and a single in-gap peak is found for
the negative impurity potential, and it shifts towards the lower
energy with the decreasing of the potential strength. These results can
be used to detect the paring symmetry of the $\mathrm{BiS_{2}}$-based superconductors.

The starting model Hamiltonian with the hopping elements, the
pairing terms and the impurity part is expressed by
\begin{equation}\label{1}
    H=H_{t}+H_{\Delta}+H_{\mathrm{imp}}.
\end{equation}
In the present work, we use the two-orbital model with the hopping
constants from Ref. 11. Thus, the hopping term $H_{t}$ can
be expressed by
\begin{eqnarray}\label{2}
    H_{t}=-\sum_{i\mu{j\nu}\sigma}(t_{i\mu{j}\nu}c_{i\mu\sigma}^{\dagger}c_{j\nu\sigma}+\mathrm{H.c.})-t_{0}\sum_{i\mu\sigma}c_{i\mu\sigma}^{\dagger}c_{i\mu\sigma},
\end{eqnarray}
where $i$,$j$ are the site indices, $\mu,\nu=1,2$ are the
orbital indices, and $t_{0}$ is the chemical potential.
Considering the possibilities of both spin singlet and triplet
pairings, the paring term is written as
\begin{eqnarray}\label{3}
    H_{\Delta}=\sum_{ij}[\Delta_{i\mu{j}\nu}^{\pm}(c_{i\mu\uparrow}^{\dagger}c_{j\nu\downarrow}^{\dagger}\pm{c_{i\mu\downarrow}^{\dagger}c_{j\nu\uparrow}^{\dagger}})+\mathrm{H.c.}].
\end{eqnarray}
Here, $\pm$ is for spin-triplet and singlet pairings respectively, and the pairing potential
$\Delta_{i\mu{j}\nu}^{\pm}$ is defined as
$\Delta_{i\mu{j}\nu}^{\pm}=\frac{V_{i\mu{j\nu}}}{2}(\langle{c_{i\mu\uparrow}c_{j\nu\downarrow}\rangle}\pm{\langle{c_{i\mu\downarrow}c_{j\nu\uparrow}}\rangle})$,
where $V_{i\mu{j\nu}}$ is the SC interaction. $H_{\mathrm{imp}}$ is the impurity part of the Hamiltonian,
written as
\begin{eqnarray}\label{4}
    H_{\mathrm{imp}}=\sum_{i_{\mathrm{m}}\mu\sigma}V_{s}c_{i_{\mathrm{m}}\mu\sigma}^{\dagger}c_{i_{\mathrm{m}}\mu\sigma}.
\end{eqnarray}
In this paper, both single- and multiple- impurity effects are
studied, and following Refs. 29, 30 and 36, only the intraorbital
scattering by nonmagnetic impurities is considered.

Then, the Hamiltonian can be diagonalized by solving the BdG
equations self-consistently,
\begin{equation}\label{5}
\sum_j \sum_\nu\left( \begin{array}{cc}
 H_{i\mu{j}\nu\sigma} & \Delta_{i\mu j\nu}^{\pm}  \\
 \mp\Delta^{\pm{*}}_{i\mu j\nu} & -H^{*}_{i\mu{j}\nu\bar{\sigma}}
\end{array}
\right) \left( \begin{array}{c}
u^{n}_{j\nu\sigma}\\v^{n}_{j\nu\bar{\sigma}}
\end{array}
\right) =E_n \left( \begin{array}{c}
u^{n}_{i\mu\sigma}\\v^{n}_{i\mu{\bar{\sigma}}}
\end{array}
\right),
\end{equation}
where the Hamiltonian $H_{i\mu{j\nu}\sigma}$ is expressed by,
\begin{eqnarray}\label{6}
     H_{i\mu{j}\nu\sigma}=-t_{i\mu{j}\nu}-t_{0}\delta_{ij}\delta_{\mu\nu}+\sum_{i_{m}}V_{s}\delta_{i,i_{m}}.
\end{eqnarray}
The SC order parameter $\Delta_{i\mu{j}\nu}$ and the local
electron density $\langle{n_{i\mu}}\rangle$ are obtained
self-consistently:
\begin{equation}\label{7}
    \Delta_{i\mu{j}\nu}^{\pm}=\frac{V_{i\mu{j}\nu}}{4}\sum_{n}(u_{i\mu\uparrow}^{n}\upsilon_{j\nu\downarrow}^{n*}\mp{u_{j\nu\uparrow}^{n}\upsilon_{i\mu\downarrow}^{n*}})\mathrm{tanh}\bigg(\frac{E_{n}}{2k_{B}T}\bigg),
\end{equation}
\begin{equation}\label{8}
    \langle{n_{i\mu}}\rangle=\sum_{n}|u_{i\mu\uparrow}^{n}|^{2}f(E_{n})+\sum_{n}|\upsilon_{i\mu\downarrow}^{n}|^{2}[1-f(E_{n})].
\end{equation}
Here $f(x)$ is the Fermi distribution function. The LDOS is
expressed by
\begin{equation}\label{9}
    \rho_{i}(\omega)=\sum_{n\mu}[|u_{i\mu\sigma}^{n}|^{2}\delta(E_{n}-\omega)+|\upsilon_{i\mu\bar{\sigma}}^{n}|^{2}\delta(E_{n}+\omega)],
\end{equation}
where the delta function $\delta(x)$ has been approximated by
$\Gamma/\pi(x^{2}+\Gamma^{2})$ with the quasiparticle damping
$\Gamma=0.002$.

Following Ref.17, we focus our studies on the optimal doped
sample with $x = 0.56$, which is used throughout the present work. The
paring strength is chosen to be $V_{i\mu{j\nu}}=0.6\mathrm{eV}$
for all paring states. The numerical calculation is performed on a
$24\times24$ lattice with the periodic boundary conditions. In
this paper, the energy and length are measured in units of
$\mathrm{eV}$ and the Bi-Bi distance $a$ respectively. To
calculate the LDOS, a $80\times80$ supercell technique is used.

To search for the possible paring symmetries, we consider onsite,
NN and NNN singlet parings as well as NNN triplet paring. The SC
order parameters are calculated self-consistently according to the
BdG technique with random initial values. For the onsite singlet
paring, the calculated SC order parameter is isotropic, so it
is a conventional s-wave state. For the singlet paring between the
NN sites, the calculated order parameter has amplitude
$+\Delta_{0}$ along the $x$ direction and $-\Delta_{0}$ along the
$y$ direction, resulting in the d-wave state with the $k$-dependent
pairing form
$\Delta(k)=\Delta_{0}(\mathrm{cos}{k_{x}}-\mathrm{cos}{k}_{y})$.
In order to study the impurity effect in the extended s-wave state
with the singlet paring between NN sites, we set the same sign along
the $x$ and $y$ directions so that the gap function has the form
$\Delta(k)=\Delta_{0}(\mathrm{cos}{k_{x}}+\mathrm{cos}{k}_{y})$.
For the singlet paring between NNN sites, the order parameter has
amplitude $+\Delta_{1}$ along the $x=y$ direction and
$-\Delta_{2}$ along the $x=-y$ direction with
$\Delta_{1}\neq\Delta_{2}$ due to the anisotropic hopping. Hence,
the gap consists of both s- and d- wave components. For the triplet
paring between NNN sites, the paring interaction is only
considered in the $x=y$ direction of the $p_{x}$ orbit and in the
$x=-y$ direction of the $p_{y}$ orbit. The calculated amplitude of
the order parameter has the form $\Delta_{i,i+x+y}=-\Delta_{i+x+y,i}$
of the $p_{x}$ orbit and $\Delta_{i,i+x-y}=-\Delta_{i+x-y,i}$ of the
$p_{y}$ orbit, resulting in the p-wave state with $k$-dependent
pairing form $\Delta(k)=\Delta_{0}\mathrm{sin}(k_{x}\mp{k_{y}})$,
where "-, +" are for the two orbits respectively.

\begin{figure}
    \includegraphics{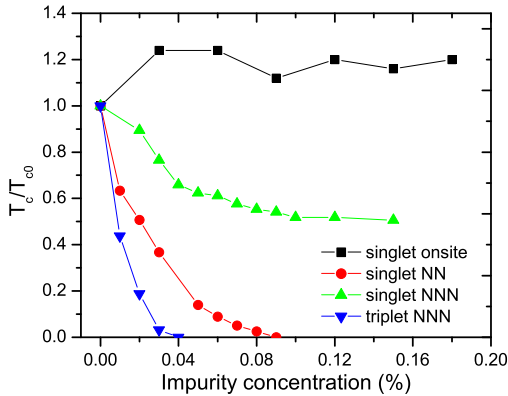}
\caption {(Color online) The impurity concentration dependence of
the reduced critical transition temperature $T_{c}/T_{c0}$ for
various paring states, where $T_{c0}$ is the critical transition
temperature without any impurity.}
\end{figure}

We first study the multiple-impurity effect for various paring
states. In this case, the impurities are randomly distributed in
the $24\times24$ lattice. The zinc element has a stable $d^{10}$
configuration in the alloy and can serve as the best non-magnetic
impurity. If the Zn impurity is located in the $\mathrm{BiS_{2}}$
layer, the scattering is quite strong. Hence, only the strong
impurity potential $V_{s}=10\mathrm{eV}$ is considered here. We
have considered 20 sets of multiple-impurity configurations and
the critical transition temperature is the average value of these
configurations. The impurity concentration dependence of the
reduced critical transition temperature $T_{c}/T_{c0}$ is
presented in Fig. 1, where $T_{c0}$ is the critical transition
temperature without any impurity. As seen, for the onsite paring
s-wave state, the critical temperature $T_{c}$ does not decrease
with the increasing impurity concentration. While, it decreases
with increasing impurity concentration for the other paring
states, such as spin singlet NN paring, NNN paring, and spin
triplet NNN paring states. For the spin singlet NNN paring state,
the critical temperature decreases rapidly in the low impurity
concentration region ($\leq5\%$), then it almost saturates with
further increasing the impurity concentration. For the spin
singlet NN paring d-wave state and the spin triplet NNN paring
p-wave state, the critical transition temperature decreases with
the increasing impurity concentration monotonically. While, the
reduction of the critical transition temperature in the p-wave state is
faster than that in the d-wave state.

Generally, studying the non-magnetic impurity effect on
superconductors helps greatly to investigate the pairing symmetry.
It is well known that superconductivity is robust against small
concentrations of non-magnetic impurities in conventional s-wave
superconductors, according to the Anderson's
theorem\cite{Anderson1959}. One possible explanation is that since superconductivity is due to the
instability of the Fermi surface to pairing of time-reversed
quasiparticle states, any perturbation that does not lift the
Kramers degeneracy of these states does not affect the mean-field
SC transition temperature. However, a magnetic impurity, owing to
the effect of breaking the time reversal symmetry, can break
Cooper pairs easily. For instance, $\mathrm{MgB_{2}}$ is a
BCS-type supercondoctor with the electron-phonon coupling and its
symmetry is a s-wave, revealed by enormous
experiments\cite{Muranaka2011}, such as the isotope effect by the
substitution of $^{11}\mathrm{B}$\cite{Bud2001} and the
observation of a coherent peak in the nuclear magnetic resonance
(NMR) experiment\cite{Kotegawa2001}. Zn ions have a full d shell
and are nominally non-magnetic. Effect of Zn-substitution on the
$\mathrm{MgB_{2}}$ superconductor shows that the critical
transition temperature $\mathrm{T_{c}}$ is almost unchanged with
small Zn-concentration, while $\mathrm{T_{c}}$ is significantly
suppressed by the substitution of magnetic ions\cite{Xu2001}.
These results are consistent with the Anderson's theorem. For the
$\mathrm{BiS_{2}}-$based superconductor, the density function
based calculations shows that this material may be a conventional
s-wave superconductor with a large electron-phonon coupling
constant\cite{Wan,Li}, but the non-magnetic impurity effect in the multi-orbital system is subtle.
For the iron-based superconductors, theoretical calculations have shown that, in the strong
scattering limit, the non-magnetic impurity effect on the $\mathrm{s_{\pm}}$-wave state is severe
and similar to the effect on the d-wave SC state\cite{Bang2009}. Experimentally, it is found that
with the presence of Zn impurity, the SC transition temperature increases in the under-doped regime,
remains unchanged in the optimally doped regime and is severely suppressed in the over-doped regime\cite{Li2010}.
The severe suppression of $\mathrm{T_{c}}$ in the over-doped regime may be well explained within the scenario of
$\mathrm{s_{\pm}}$-wave symmetry. On the other hand, the insensitivity of the impurity effect in the under-doped
and optimally doped regimes is not in accordance with the $\mathrm{s_{\pm}}$ pairing but with the s-wave state
corresponding to the same signs of the relative order parameters between the hole and electron Fermi pockets.
A very recent theoretical calculation based on a two-orbital model for the iron-based superconductors also indicates
that $\mathrm{T_{c}}$ is insensitive to the Zn impurity if the SC order parameter has a large s-wave component with
the relatively strong onsite paring strength\cite{Yao2012}. For the $\mathrm{BiS_{2}}$-base superconductor,
the symmetry of the order parameter is an isotropic s-wave state if considering the onsite paring.
As revealed by our numerical calculation presented in Fig. 1, $\mathrm{T_{c}}$ is not sensitive to
the non-magnetic impurities in the s-wave state, which is consistent with the results in the iron-based
compounds discussed above. Note that there is an
enhancement of $\mathrm{T_{c}}$ by the non-magnetic impurity
substitution, which is perhaps due to the enhanced density of
state near the Fermi level\cite{Kemper2009,Garg2008}.

In contrast to the isotropic s-wave superconductors, non-magnetic
impurities are paring breakers in anisotropic superconductors,
such as heavy fermion superconductors\cite{Millis1988} and
cuprats\cite{Radtke1993}. Based on the second Born approximation
and the strong-coupling Eliashberg theory, impurity-induced
$\mathrm{T_{c}}$ suppression in d-wave cuprate superconductors has
been predicted\cite{Radtke1993}. Experimentally, Zn substitution
for Cu in the high-$\mathrm{T_{c}}$ cuprates dramatically
suppresses superconductivity. For instance, it has been shown that
an impurity concentration of 2-3 at.$\%$ (per Cu) reduces
$\mathrm{T_{c}}$ to half or less for unsubstituted
systems over a wide region of hole doping level in Zn substituted
$\mathrm{La_{2-y}Sr_{y}CuO_{4}}$ and
$\mathrm{YBa_{2}Cu_{3}O_{6+\delta}}$\cite{Xiao1990,Fukuzumi1996,Mendels1994}.
In our self-consistent calculation, it is found that the paring
symmetry is d-wave in the NN singlet paring state for
$\mathrm{BiS_{2}}-$compounds. The critical transition temperature
$\mathrm{T_{c}}$ is suppressed significantly by small
concentration of non-magnetic impurities. This result is similar to
that of high $\mathrm{T_{c}}$ cuprate superconductors, indicating
the paring breaking effect in d-wave SC gap. While in the NNN
singlet paring state, $\mathrm{T_{c}}$ is suppressed quickly for
small impurity concentration then saturates with further
increasing impurities. According to our calculation, the SC gap
has both s- and d- wave components. Hence, the suppression of
$\mathrm{T_{c}}$ is the consequence of the paring break effect of
the d-wave component in the SC gap, but the saturation of
$\mathrm{T_{c}}$ is due to the s-wave component.

In principle, non-magnetic impurities are pair breakers for
higher-orbital momentum states like p-wave
state\cite{Balatsky2006}. The reduction of $\mathrm{T_{c}}$ due to
non-magnetic impurity scattering in $\mathrm{Sr_{2}RuO_{4}}$
compound with p-wave like symmetry has been calculated with the
conventional Abrikosov-Go\'{r}kov formula\cite{Miyake1999}, which
has been proved by the substitution of non-magnetic impurity
$\mathrm{Ti^{4+}}$ ions\cite{Kikugawa2002,Minakata2001}. For the
$\mathrm{BiS_{2}}-$based superconductors, possible spin triplet
p-wave paring state is proposed\cite{Usui,Zhou}. From our
self-consistent calculation, the rapid depression of
$\mathrm{T_{c}}$ by the non-magnetic impurity substitution can be
tested experimentally.

\begin{figure}
    \includegraphics{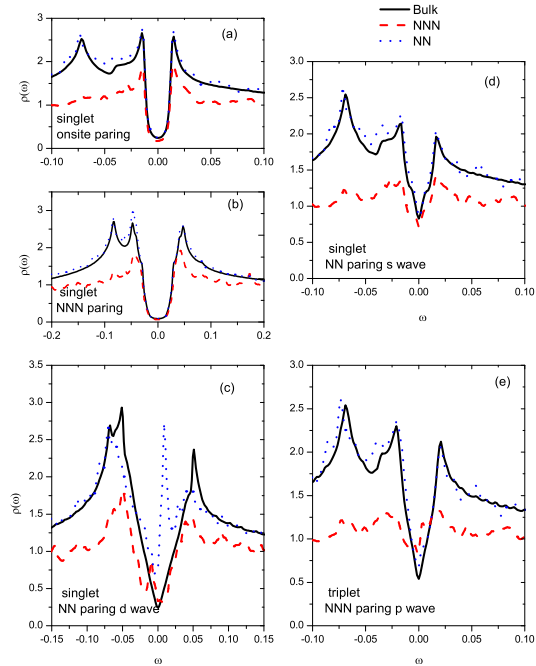}
\caption {(Color online) LDOS spectra around a non-magnetic
impurity at various paring states. The impurity strength is set to
be $V_{s}=10\mathrm{eV}$. The solid (black) curves are the LDOS
spectra in the bulk, the dash dotted (green) curves are the LDOS
spectra at the NN sites of the impurity, and the dashed (red)
curves are the LDOS spectra at the NNN sites of the impurity. (a):
Spin singlet of onsite paring state; (b): Spin singlet of NNN
paring state; (c): Spin singlet of NN paring for d wave state;
(d): Spin singlet of NN paring for extended s wave state; (e):
Spin triplet of NNN paring for p wave state.}
\end{figure}

We now turn to study the single impurity effect in the LDOS
spectra of various paring symmetries. The calculated results are
presented in Fig. 2(a)-2(e). Shown in Fig. 2(a) is the LDOS
spectra around a non-magnetic impurity for the isotropic s-wave
paring state. As it can be seen, the LDOS spectrum in the bulk has a
U-shaped bottom, indicating a full SC gap. Outside the gap, the
van Hove singularity peak is clearly seen. At the NN site of the
impurity, the LDOS spectrum is almost unchanged compared with that
in the bulk, while at the NNN site of the impurity, the SC
coherence peaks in the LDOS spectrum are depressed. From the above
result, it is found that there is no impurity state in the
isotropic s-wave paring state. In general, the potential scattering
impurities are not pair breakers in the s-wave case according to
the Anderson theorem\cite{Balatsky2006}. Presented in Fig. 1(b) is
the LDOS spectra for the spin singlet NNN paring state. The
results are similar to the case of the onsite paring state, i.e.,
the LDOS spectrum has a U-shaped bottom and no impurity state
occurs inside the SC gap.

The LDOS spectra for the spin singlet NN paring d-wave state are
shown in Fig. 2(c). It is found that the LDOS spectrum has a
V-shaped bottom in the bulk, indicating the nodal characteristics.
At the NN site of the impurity, there is a sharp resonance peak at
$\omega_{0}=9\mathrm{meV}$ in the LDOS spectrum. At the NNN site
of the impurity, there is also a weak in-gap peak at
$\omega_{0}=-9\mathrm{meV}$. Note that, the SC coherence peaks are
also depressed obviously by the impurity at the NNN site. The
resonance peak inside the SC gap at the NN site of the impurity is
a general feature in a d-wave superconductor, which has already
been reported in cuprate superconductors both
theoretically\cite{Balatsky2006} and experimentally\cite{Pan2000}.
The LDOS spectra for the spin singlet NN paring s-wave state are
presented in the Fig. 2(d). The LDOS spectrum in the bulk still
has a V-shaped bottom. The LDOS spectrum at the NN site of the
impurity is almost unchanged compared with that in the bulk.
Similar to the spin singlet onsite and NNN paring states, there is
no impurity state in this case. However, at the NNN site of the
impurity, the SC coherence peaks are almost flattened, which is
quite different from those in the previous cases.

The LDOS spectra for the spin triplet NNN paring p-wave state are
shown in Fig. 2(e). As one can see, the LDOS spectrum in the bulk
is also V-shaped with nodal characteristics. Previously, a fully
nodeless SC gap was found in the $\mathrm{Sr_{2}RuO_{4}}$
superconductors from the scanning tunnelling microscopy
experiments\cite{Suderow2009}, which is quite different from our
result. This is most due to its complex paring symmetry without
nodes at the Fermi surface, i.e.,
$\mathrm{sin}{p_{x}}+i\mathrm{sin}{p_{y}}$ wave or
$\mathrm{sin}({p_{x}}+p_{y})+i\mathrm{sin}(-p_{x}+{p_{y}})$
wave\cite{Takigawa2005}. At the NN site of the impurity, there is
a small in-gap peak at $\omega_{0}=3\mathrm{meV}$ in the spectrum.
At the NNN site of the impurity, two in-gap peaks are found at
$\omega_{0}=\pm3\mathrm{meV}$, where the in-gap peak at the
positive energy is much stronger than that at the negative energy.
Moreover, the SC coherence peaks are almost flattened by the
impurity, similar to that of the spin singlet NN paring s-wave
state. From the above discussion, one can conclude that there are
impurity resonance peaks only in the spin singlet NN paring d-wave state.
The resonance peak is very sharp in the LDOS spectrum
at the NN site. However, two in-gap peaks occur in the spectrum at
the NNN site for the p-wave paring state and the SC coherence
peaks are almost flattened by the impurity. These results can be
used to detect the paring symmetry in the $\mathrm{BiS_{2}}$-based superconductors.

\begin{figure}
    \includegraphics{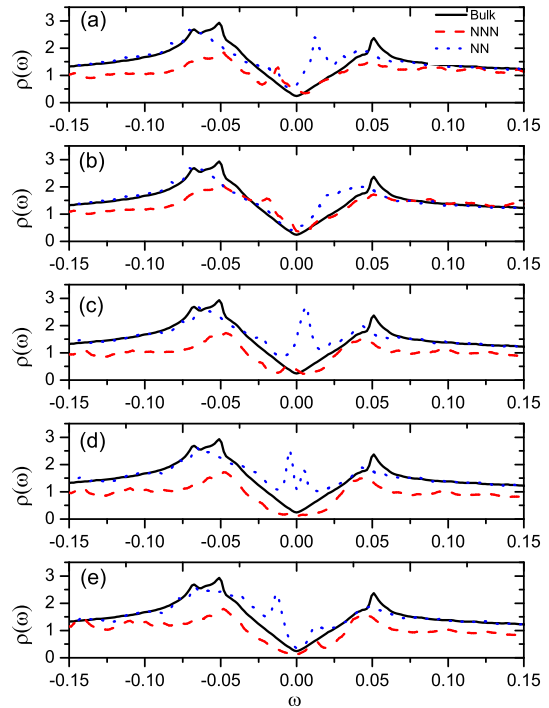}
\caption {(Color online) LDOS spectra around a non-magnetic
impurity at various potential strengths for d-wave paring state.
(a): $V_{s}=3\mathrm{eV}$; (b): $V_{s}=1\mathrm{eV}$; (c):
$V_{s}=-10\mathrm{eV}$; (d): $V_{s}=-2\mathrm{eV}$;
(e):$V_{s}=-1\mathrm{eV}$.}
\end{figure}

\begin{figure}
    \includegraphics{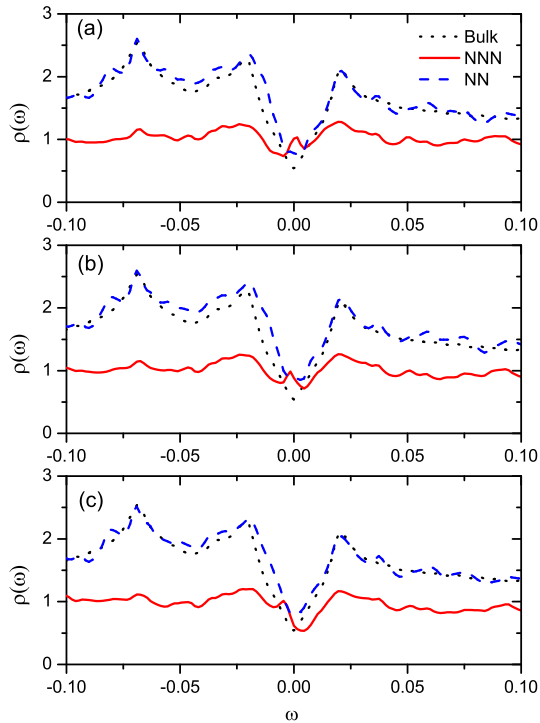}
\caption {(Color online) LDOS spectra around a non-magnetic
impurity at various potential strengths for p-wave paring state.
(a): $V_{s}=-10\mathrm{eV}$; (b): $V_{s}=-3\mathrm{eV}$; (c):
$V_{s}=-1.5\mathrm{eV}$.}
\end{figure}

The potential strength response of the impurity state is also
calculated. The impurity states at various potential strengths for
the d-wave paring state are presented in Fig. 3(a)-3(e). As seen
in Fig. 3(a)-3(b), for the positive impurity potential the
resonance peak at the NN site of the impurity decreases with the
decreasing of the potential strength. At the impurity potential
$V_{s}=1\mathrm{eV}$, the resonance peak almost disappears and
merges into the SC coherence peak. However, the situation is quite
different in the case of the negative impurity potential. At the
strong impurity potential $V_{s}=-10\mathrm{eV}$, there is also a
sharp resonance peak at the NN site of the impurity located at
$\omega_{0}=6\mathrm{meV}$, which is shifted to the lower energy
compared with that of the impurity potential
$V_{s}=10\mathrm{eV}$. With the impurity potential decreasing, as
shown in Fig. 3(d), the resonance peak at the NN site of the
impurity splits into two peaks located at
$\omega_{0}=\pm4.5\mathrm{meV}$, and the peak at the negative
energy is much stronger than that at the positive energy. With the
impurity potential $V_{s}=-1\mathrm{eV}$, there are also two
in-gap peaks at the NN site of the impurity located at
$\omega_{0}=\pm13.5\mathrm{meV}$, and the peak at the positive
energy almost disappears.

The potential strength response of the impurity states for the
p-wave paring state is presented in Fig. 4(a)-4(c). For the
positive impurity strength, the LDOS spectra are similar to that
of $V_{s}=10\mathrm{eV}$, i.e., two in-gap peaks occurring at the
NNN site of the impurity. These results are not presented for
simplicity. Here, we focus on the LDOS spectra of the negative
impurity potential for the p-wave paring state. As it can be seen, at
the NN site of the impurity, the LDOS spectra are almost
featureless compared with those in the bulk. However, there is an
in-gap peak in the LDOS spectra at the NNN site of the impurity,
and the SC coherence peaks are almost flattened. Moreover, the
location of the in-gap peak shifts from
$\omega_{0}=1.5\mathrm{meV}$ at $V_{s}=-10\mathrm{eV}$ to
$\omega_{0}=-4.5\mathrm{meV}$ at $V_{s}=-1.5\mathrm{eV}$.

In summary, the impurity effects are studied for the
$\mathrm{BiS_{2}}$-layered superconductors based on a two-orbital
model with the BdG technique. From the calculation of multiple-impurity effect,
the significant reduction of $\mathrm{T_{c}}$ is
found for the spin singlet NN paring d-wave state and the spin triplet NNN
paring p-wave state, while no depression of $\mathrm{T_{c}}$ for
isotropic s-wave state. The single-impurity effects for various
paring states are also calculated. The impurity resonance peak in
the LDOS spectrum is found only for the d-wave state. For the spin
triplet NNN paring p-wave state, two in-gap peaks occur in the
case of positive impurity potential, while a single in-gap peak is
found for the negative impurity potential, and it shifts towards
the lower energy with the decreasing of the potential strength. These
results can be used to detect the paring symmetry of the
$\mathrm{BiS_{2}}$-based superconductors.

This work is supported by the research foundation of the Nanjing
university of posts and telecommunications (NY211146). We also
appreciates the mathematical laboratory of the college of science,
Nanjing university of posts and telecommunications for the
technical support.

\end{document}